\documentclass[11pt]{article}
\usepackage{amsmath}
\usepackage{epsf,afterpage,epsfig}

\def\abstracts#1#2#3{{
        \centering{\begin{minipage}{4.62in}\baselineskip=13pt
        \small
        \centerline{\bf Abstract}
        \vspace*{0.2cm}                
        \parindent=0pt #1\par
        \parindent=18pt #2\par
        \parindent=15pt #3
        \end{minipage} }\par}}
\begin{document}
\vspace*{-2cm}
\hfill \parbox{4cm}{ ~\\~}\\
%
\centerline{\LARGE \bf Ising and Potts Models on}\\[0.3cm]
\centerline{\LARGE \bf Quenched Random Gravity Graphs}\\[0.4cm]
\vspace*{0.2cm}
\centerline{\large {\em Wolfhard Janke$^1$ and 
Desmond A. Johnston$^2$\/}}\\[0.4cm]
\centerline{\large {\small $^1$ Institut f\"ur Theoretische Physik,
                    Universit\"at Leipzig}}
\centerline{    {\small D-04109 Leipzig, Germany }}\\[0.5cm]
\centerline{\large {\small $^2$ Department of Mathematics, 
                    Heriot-Watt University}} 
\centerline{    {\small Edinburgh, EH14\,4AS, Scotland  }}\\[0.5cm]
\abstracts{}{We report on
single-cluster Monte Carlo simulations
of the Ising, 4-state Potts and 10-state Potts models
on quenched ensembles 
of planar, tri-valent ($\Phi^3$) random graphs. We confirm
that the first-order phase transition of the 10-state Potts model 
on regular 2D lattices is softened by the quenched connectivity disorder 
represented 
by the random graphs and that the exponents
of the Ising and 4-state Potts models are altered from
their regular lattice counterparts. The behaviour of spin models on such graphs is thus
more analogous to models with quenched bond disorder than to 
Poisonnian random lattices, where regular lattice critical behaviour persists.  
\\
\indent
Using a wide variety of estimators we measure 
the critical exponents for all three models, 
and compare the exponents 
with predictions derived from taking a quenched limit in the KPZ formula
for the Ising and 4-state Potts models. 
Earlier simulations
suggested that the measured values for the 10-state Potts model
were very close to the predicted quenched exponents of the {\it four}-state 
Potts models. The analysis here,
which employs a much greater range of estimators and also benefits
from greatly improved statistics, still supports
these numerical values.
}{}
%
\thispagestyle{empty}
\newpage
\pagenumbering{arabic}
%
                     \section{Introduction}
%
There has recently been some interest, and no little controversy, regarding the
critical behaviour of systems with quenched bond disorder in 2D \cite{1}.
It has been known for some time that the first-order phase transition displayed by $q>4$-state Potts models on regular
lattices is softened by the introduction of the quenched bond disorder to a continuous transition
\cite{2},
though the universality class of this transition and its dependence on the strength and nature of
the bond disorder are still not completely clear \cite{3,4}. Models which already display a 
continuous transition in the pure case appear to have their critical exponents altered
by the bond disorder \cite{5,5a,5b} provided that the critical exponent 
$\alpha_{\rm pure}$
of the
specific heat is positive. 
In addition, a qualitatively new phenomenon in the form
of multi-fractal scaling of local correlators has also been predicted \cite{6} and measured \cite{6a}
when quenched bond disorder is present. It should be remarked that all of the theoretical
results in the bond disordered case are perturbative in nature, which is in large measure
the source of the controversy surrounding the various predictions for critical exponents,
since the domain of validity is unclear.

Another type of disorder that might be imposed is quenched
{\it connectivity} disorder. Very high accuracy numerical 
simulations have shown that spin models on Poisonnian random lattices
in both 2D \cite{7} and 3D \cite{8} stay stubbornly identical to their regular lattice brethren
- there is no sign of the effects observed with quenched bond disorder.
However, a different picture emerges when one considers spin models
living on a quenched ensemble
 of tri-valent ($\Phi^3$) planar graphs, as generated
by simulations of 2D quantum gravity. In this case the connectivity disorder
is sufficiently strong for  $q>4$ Potts model transitions
to be softened to continuous transitions \cite{9} 
and for $q \le 4$ exponents to be modified from their regular 2D lattice values \cite{10}.
In this respect such planar random graphs appear to be much more akin to random bond disorder
models than to the Poisonnian random lattices
considered in \cite{7,8}. One very interesting feature
of the $\Phi^3$ graphs is that exact, rather than perturbative, predictions for exponents
exist in the $q \le 4$ case, by virtue of taking a quenched limit \cite{6,11} in the KPZ
formula which gives the weights of conformal operators when they are
coupled to 2D quantum gravity.

In the quantum gravity and string theory context one is typically interested in
{\em annealed} rather than quenched connectivity disorder, 
in which the lattices and spins are interacting on the same time scale,
providing a discrete analogue of the back reaction in continuum
theories of gravity. In this case  
the relation between the bare ($\Delta$) and dressed ($\tilde \Delta$)
weights is given by the KPZ relation \cite{12}
\begin{equation}
\tilde \Delta = { \sqrt{1 - c + 24 \Delta} - \sqrt{1 -c } \over \sqrt{25 -c} - \sqrt{1 -c} }.
\end{equation}
In order to calculate the dressed weights in the quenched case one should take $c=0$
in the KPZ relation to get
\begin{equation}
\tilde \Delta_{quenched} = { \sqrt{1 + 24 \Delta} - {1 } \over 4 },
\end{equation}
which now gives non-rational weights. Indeed, Cardy \cite{6} has recently pointed out that
the $n$th power of a correlator
with bare weight $\Delta$ averaged over the disorder
will scale not as $n \tilde \Delta$, but rather
\begin{equation}
\tilde \Delta_n = { \sqrt{1 + 24 n \Delta} - {1 } \over 4 },
\end{equation}
with a ``typical'' \footnote{Rather than average.}
value being governed by \cite{6}
\begin{equation}
\left. { \partial \tilde \Delta_n  \over  \partial n} \right|_{n=0}   = 3 \Delta.
\end{equation}

In this paper we concern ourselves exclusively with 
measuring quantities which correspond to
$n=1$ in the formula above,
such as the specific heat, susceptibility
and magnetisation. The picture suggested by the earlier $q=2$ and $q=10$ simulations
in \cite{9,10} was that 
measurements were in accordance with the predicted quenched exponents
for $q \le 4$
and that the exponents of the quenched $q > 4$ models
were ``stuck'' at the $q=4$ values, as evinced by
the good agreement between the measured $q=10$ exponents and the
quenched $q=4$ predictions. This was somewhat similar to the original scenario
postulated for the quenched bond disordered Potts models,
where it was suggested that all bond disordered $q>4$ Potts models
displayed Ising-like criticality. An obvious test of the scenario
for quenched connectivity disorder 
is to perform simulations at $q=4$, as well
as improving the quality of the measurements for $q=2$ and $q=10$ in order
to get sharper estimates of the exponents in these cases. 
This is precisely what we do here.

In what follows we briefly describe the simulation methods, an
extension of those used in \cite{9}, before going on to discuss
the analysis of the results and our best estimates for the various exponents
for the Ising, 4-state Potts and 10-state Potts models. We finish
with some observations on the values we obtain.
%
                     \section{Simulation and Measurements}
%
We used the standard definition of the $q$-state Potts model
partition function and energy in all the simulations,
\begin{equation}
Z_{\rm Potts} = \sum_{\{\sigma_i\}} e^{-\beta E}; \; \; 
E = -\sum_{\langle ij \rangle}
\delta_{\sigma_i \sigma_j};  \; \; \sigma_i = 1,\dots,q,
\label{eq:zpotts}
\end{equation}
where $\beta=J/k_B T$ is the inverse temperature in natural units, 
$\delta$ is the Kronecker symbol, and 
$\langle ij \rangle$ denotes the nearest-neighbour bonds of random
$\Phi^3$ graphs (without tadpoles or self-energy bubbles) with 
$N$ sites. In this study we considered the cases $q=2$ and $4$
(with $N = 500$, 
1\,000, 2\,000, 3\,000, 4\,000, 5\,000, and 10\,000) which in the
pure model exhibit second-order phase transitions, and the case
$q=10$ (with $N = 250$, 500, 1\,000, 2\,000, 3\,000, 5\,000, 
and 10\,000) which in the pure model undergoes a first-order
phase transition.

The simulations were carried out using the Wolff single-cluster update 
algorithm \cite{13}.
For each
lattice size we generated 64 independent graphs using the Tutte algorithm
\cite{14}, and performed 500K equilibration sweeps followed by
up to 10 million measurement sweeps  in order to obtain 500K independent
measurement sweeps for each lattice size.
The runs were carried out at several
$\beta$ values near the transition point 
and time series of the energy $E$ and the magnetisation%
\footnote{Where $n_i \le N$ denotes the number of spins of ``orientation''
$i=1,\dots,q$ in one lattice configuration.} 
$M = (q \, {\rm max} \{ n_i \} - N)/(q-1)$ 
recorded for each graph.
In what follows the per-site quantities are denoted by $e = E/N$ and $m = M/N$,
the thermal averages on each individual graph by $\langle \ldots \rangle$ and the
quenched average over the different graphs by $[ \ldots ]_{\rm av}$.

From the time series of $e$ and $m$ it is straightforward to
compute in the finite-size scaling (FSS) region various quantities at 
nearby values of $\beta$  by
standard reweighting \cite{15} techniques.
Some care must be taken with the reweighting range in the presence
of quenched averaging, but we confirmed that direct measurements
of both the susceptibility and specific heat 
from fluctuations and numerical derivatives were in accordance with
the values deduced from reweighting in several representative cases.
Comparisons for $q=10$ and $N=2000$ around the reweighting
point of $\beta_0=2.22$ are shown in Figs.~1 and 2.

To estimate the statistical (thermal) errors for each of the 64 realizations,
the time-series data was split into
bins, which were jack-knifed \cite{16} to decrease the bias in the
analysis of reweighted data.
The final values are averages over the 64 realizations which will be denoted
by square brackets $[\dots]_{\rm av}$, and the error bars are computed from the
fluctuations among the realizations. Note that these errors contain both
the average thermal error for a given realization and the theoretical variance
for infinitely accurate thermal averages which is caused by the variation
over the random graphs.

From the time series of the energy measurements we compute by reweighting the
average energy, specific heat, and energetic fourth-order cumulant,
\begin{eqnarray}
u(\beta) &=& [\langle E \rangle]_{\rm av}/N, \nonumber \\
C(\beta) &=& \beta^2 \, N [\langle e^2 \rangle - \langle e \rangle^2]_{\rm av},\\
V(\beta) &=&  [1 - \frac{\langle e^4 \rangle}{3 \langle e^2 \rangle^2}]_{\rm av} \nonumber .
\label{eq:ene}
\end{eqnarray}
Similarly, we derive from the magnetisation measurements the average
magnetisation, susceptibility, and magnetic cumulants
\footnote{See below for some subtleties in the ordering of averages
in the cumulants.},
\begin{eqnarray}
m(\beta) &=& [\langle |m| \rangle]_{\rm av}, \nonumber \\
\chi(\beta) &=& \beta \, N [ \langle m^2 \rangle - \langle |m| \rangle^2 ]_{\rm av}, \nonumber \\
U_2(\beta) &=& [1 - \frac{\langle m^2 \rangle}{3 \langle |m|
\rangle^2}]_{\rm av},\\
U_4(\beta) &=& [1 - \frac{\langle m^4 \rangle}{3 \langle m^2 \rangle^2}]_{\rm av}.
\label{eq:mag} \nonumber 
\end{eqnarray}
Further useful mixed quantities involving both the energy and magnetisation are
defined by 
\begin{eqnarray}
\frac{d [\langle |m| \rangle]_{\rm av}} {d \beta} &=&
  [\langle |m| E \rangle  - \langle |m| \rangle \langle E \rangle]_{\rm av}, \nonumber \\
\frac{d \ln [\langle |m| \rangle]_{\rm av}} {d \beta} &=&
\left[ \frac{\langle |m| E \rangle}{\langle |m| \rangle} - \langle E
\rangle\right]_{\rm av},\\
\frac{d \ln [\langle m^2 \rangle]_{\rm av}} {d \beta} &=&
\left[ \frac{\langle m^2 E \rangle}{\langle m^2 \rangle} - \langle E
\rangle \right]_{\rm av} \nonumber .
\end{eqnarray}

The dynamical aspects of the simulations are encoded in the autocorrelation
functions and the associated integrated autocorrelation times $\hat{\tau}$.
It is customary \cite{1c}
to convert the $\hat{\tau}$ thus obtained by multiplying with a factor
$f = n_{\rm flip} \langle |C| \rangle/N$ to a scale where, on the average,
measurements are taken after every spin has been flipped once (similar to,
e.g., Metropolis simulations). For quenched, random systems this procedure
is not unique due to the necessary average over realizations, since one can take
either  $[\tau]_{\rm av} \equiv
[f \cdot \hat{\tau}]_{\rm av}$ or
$[f]_{\rm av} \cdot [\hat{\tau}]_{\rm av}$. The differences between the
two averaging prescriptions turn out, however, to be extremely small in practice.

One finds that the autocorrelation times for $q=2$ stay roughly constant with increasing system
size, being $ [ \tau_e ]_{\rm av} \sim 3 - 4$ for the energy and $[ \tau_m ]_{\rm av} \sim 1.6 - 2.2$ for the magnetisation.
For $q=4$ scaling behaviour is visible with $[ \tau_e  ]_{\rm av} \sim 12 - 18$ and $[ \tau_m  ]_{\rm av} \sim 7 - 10$,
giving a dynamical exponent $z/D \sim 0.064(10)$ for the energy. Power law scaling is
much more pronounced for the   
$q=10$ model with $[ \tau_e  ]_{\rm av} \sim 60 - 500$ and $[ \tau_m  ]_{\rm av} \sim 40  - 350$  
and much larger dynamical exponents for both the energy and magnetisation, $z/D \sim O(1)$. 

The self-averaging properties of the ensemble
can be investigated by considering the probability density for the $\tau$'s, $P (\tau)$,
rather than the average values, $[ \tau_{e,m} ]_{\rm av}$. One would expect the 
cumulative distribution $F ( \tau ) = \int_0^{\tau} P ( \tau' ) d \tau'$ to tend to a step function for increasing
system size in a self-averaging system. This is observed {\it not} to be the case for all the models
simulated, giving clear evidence of non-self-averaging behaviour. These observations can be put on a 
more quantitative basis by looking at data collapse with the scaled variable $\Delta \tau / [ \tau ]_{\rm av}$
\cite{jj}, where $\Delta \tau$ is the standard deviation. 


                  \section{Data Analysis and Results}


In the infinite-volume limit the various measured quantities exhibit singularities at
the transition point. In finite systems the singularities are smeared
out and scale in the critical region according to
\begin{eqnarray}
C &=& C_{\rm reg} + N^{\alpha/\nu D} f_C(x) [1 + \dots],
\label{eq:fss_C} \nonumber \\
\chi &=& N^{\gamma/\nu D} f_\chi(x) [1 + \dots],
\label{eq:fss_chi} \nonumber \\
 {[} \langle |m| \rangle ]_{\rm av} &=&
N^{-\beta/\nu D} f_{m}(x) [1 + \dots],
\label{eq:fss_m}\\
\frac{d [\langle |m| \rangle]_{\rm av}} {d \beta} &=&
N^{(1-\beta)/\nu D} f_{m'}(x) [1 + \dots],
\label{eq:fss_dmdk} \nonumber \\
\frac{d \ln [\langle |m|^p \rangle]_{\rm av}} {d \beta} &=&
N^{1/\nu D} f_p(x) [1 + \dots],
\label{eq:fss_dlnmdk} \nonumber \\
\frac{d U_p}{d \beta} &=& N^{1/\nu D} f_{U_{2p}}(x) [1 + \dots],
\label{eq:fss_dUdk} \nonumber 
\end{eqnarray}                        
where $C_{\rm reg}$ is a regular background term, $\nu$, $\alpha$,
$\beta$, and $\gamma$, are the usual critical exponents, and the
$f_i(x)$ are various FSS functions with
\begin{equation}
x = (\beta- \beta_c) N^{1/\nu D}
\label{eq:x}
\end{equation}
being the scaling variable. $[1 + \dots]$ indicates correction terms
which become unimportant for sufficiently large system sizes $N$.
We have expressed the scaling relations in terms of the total number of vertices
$N$ rather than the linear size $L$ since the fractal dimension $D$ of the
graphs is {\it a priori} unknown. Numerical simulations and various analytic
approaches suggest that $D=4$ \cite{17} for the ensemble of graphs we are considering,
but we shall not need this explicitly for our analysis here.
By rearranging equ.(\ref{eq:x}) one finally obtains the standard scaling relation
for the peak-locations (pseudo-critical points) $\beta_c(N)$ on finite graphs,
\begin{equation}
\beta_c(N) = \beta_c + a N^{-1/\nu D},
\label{eq:bc}
\end{equation}
with $a$ being a constant.

One further issue of principle remains. In the presence of quenched disorder
there are three equally plausible ways of defining the various cumulants.
For instance, if we take the fourth order magnetic cumulant as an example we could define not only the form used above
\begin{eqnarray}
U_4^{(1)} &=& \left[ 1 -  { \langle m^4 \rangle  \over 3  \langle m^2 \rangle^2 } \right]_{\rm av}, 
\label{eq:U4_1}
\end{eqnarray}
but also the variants
\begin{eqnarray}
U_4^{(2)} &=& 1 - { \left[  \langle m^4 \rangle  \right]_{\rm av} \over 3  \left[ \langle m^2 \rangle^2  \right]_{\rm av}}, \nonumber \\
U_4^{(3)} &=& 1 - { \left[  \langle m^4 \rangle  \right]_{\rm av} \over 3  \left[ \langle m^2 \rangle  \right]_{\rm av}^2}, 
\label{eq:U4_23}
\end{eqnarray}
and the correct choice is not 
immediately clear. We can hedge our bets in the scaling analysis 
by including all three of the variants without prejudice in order 
to check their consistency.
In simulations with poor statistics per realisation (such as typically in
spin-glass studies) usually $U_4^{(3)}$ is taken since with that choice
(systematic) bias effects are minimised. In our case, however, the statistics
for each realisation is so large that there is no reason to favour
one definition over the other on technical grounds.

Without further ado, we start the analysis by estimating the 
value of $1 / \nu D$.
With the wealth of available estimators we
have various tactics available for the extraction of $1 / \nu D$.
One possibility would be to use the
maxima of
each of $dU_2/d \beta$, $dU_4/d \beta$ (all variants for both cumulants),
$d \ln [ \langle |m| \rangle]_{\rm av}/d \beta$, and
$d \ln [ \langle m^2 \rangle]_{\rm av}/d \beta$
as pseudo-critical points and then evaluate
the scaling of each of these quantities
at their own maxima to extract the exponent.
Another would be to evaluate
the scaling of each of these quantities at {\it all}
of the available pseudo-critical points.
A  global estimate is
then extracted by performing a direct
or error weighted average.                                
In both these cases
we take a fairly conservative estimate
for the errors by using
the smallest contributing error bar
\footnote{The {\it largest} contributing error bar
would certainly be too pessimistic; this choice probably errs on the side of caution too.}. 

We present the results from both approaches in Table~1.
In obtaining these estimates we have dropped the smallest
graph sizes in all cases
and used $U^{(1)}$ for definiteness as our definition
of the cumulants for the results presented in the table:  $U^{(2)}$ and  $U^{(3)}$
give values that are indistinguishable within the error
bars. The results are quite stable to the deletion of the
next smallest size, but the quality of the fits declines somewhat when this is done.
In all of the listed fits the quality of fit $Q$ was very good, the lowest being $\approx 0.3$,
with most being as high as $0.8-0.9$.
For comparison we have included the prediction of the quenched KPZ
formula (equ.(2)), the standard KPZ
exponents and the regular 2D lattice exponents
in the lower box. Since the $q=10$ model
has a first-order transition on a regular 2D lattice there
is no direct prediction in this case.
%
%
%
\begin{table}[t]
\centering
\caption[{\em Nothing.}]
 {{\em Fit results for the critical exponent $1/\nu D$.}}
\vspace{3ex}
\begin{tabular}{|l|l|l|l|l|}
\hline\hline
\multicolumn{1}{|c|}{quantity}        &  
\multicolumn{1}{c|}{type}         &
\multicolumn{1}{c|}{$q=2$}      & 
\multicolumn{1}{c|}{$q=4$}      & 
\multicolumn{1}{c|}{$q=10$}      \\
\hline 
$d U_2/d \beta$ & at maximum   &  0.32(1) & 0.42(2)& 0.59(3)\\
           & average      &  0.32(1) & 0.45(2)& 0.61(3)\\
           & weighted av. &  0.31(1) & 0.44(2)& 0.60(3)\\
\hline 
$d U_4/d \beta$ & at maximum   &  0.32(2)& 0.39(2)& 0.62(3)\\
           & average      &  0.31(1)& 0.40(2)& 0.59(2)\\
           & weighted av. &  0.31(1)& 0.40(2)& 0.59(2)\\
\hline\hline
$d \ln [\langle |m| \rangle]_{\rm av}/d \beta$
           & at maximum   &  0.36(1) & 0.42(1)& 0.56(1)\\
           & average      &  0.36(1) & 0.43(1)& 0.56(1)\\
           & weighted av. &  0.36(1) & 0.43(1)& 0.56(1)\\
\hline 
$d \ln [\langle m^2 \rangle]_{\rm av}/d \beta$
           & at maximum   &  0.36(1) & 0.42(1)& 0.56(1)\\
           & average      &  0.36(1) & 0.43(1)& 0.57(1)\\
           & weighted av. &  0.36(1) & 0.43(1)& 0.57(1)\\
\hline\hline
{\bf meta-average} &   {}     &  {\bf 0.34(1)} & {\bf 0.42(1)} & {\bf 0.58(1)} \\
\hline\hline                                                     
$Quenched$ &   {}         &  0.3486\dots & 0.589\dots& ------\\
\hline
$KPZ$      &   {}         &  0.3333\dots & 0.5 & ------\\
\hline
$Regular$  &   {}         &  0.5      & 0.75 & ------\\
\hline\hline
\end{tabular}
\label{tab:nu_inv}
\end{table}
\newpage

Looking at the results in Table~1 it is clear that the estimates of $1 / \nu D$
are not
consistent with those for regular 2D lattices, giving a clear
indication that the planar random graphs are different in this respect
from Poisonnian random lattices.
The average of averages or
``meta-average'' for $q=2$ is compatible with both the quenched and KPZ
values at the level of accuracy we have achieved, but that for $q=4$ matches
none of the possible theoretical predictions. Remarkably, the estimated $q=10$ values
are a good fit to the quenched $q=4$ predictions, as we have
already noted in \cite{9}, and
the numerous additional
estimators here add extra weight to this observation.
The difference between the theoretical quenched and KPZ values for $q=4$
is sufficient for the $q=10$ estimates {\it not} to be consistent with the
theoretical $q=4$ KPZ value.                                                  

It is also noteworthy that the $q=10$ measurements (and also
the $q=4$ quenched theory predictions) violate a supposedly general
bound derived by Chayes {\em et al.\/} \cite{18a} for quenched systems,
$\nu D > 2$, 
since $\nu D \sim 1.72(3)$ from the $q=10$ measurements. Hyperscaling,
$\alpha / \nu D = 2 / \nu D  - 1 $, implies that $\alpha / \nu D $ should be negative if the bound holds,
but we find below that direct fits to the specific heat for $q=10$ also give a positive
value ($0.21(1)$) that is compatible with that deduced from hyperscaling ($0.16(1)$).
The measured values of $\nu D$ for $q=2$ and $q=4$, on the other hand, {\it are} consistent with the bound.
Whether the failure of the $q=10$ model to observe the bound is a consequence of the technical details
of the averaging preocedure as suggested in \cite{18b} or a result of long range correlations in the disorder
(which is due to the curvature correlations for the Liouville action in
the original 2D gravity theory used to generate the graphs)
is unclear.

We now use our best estimates of $1 / \nu D$ to extract                  
the critical coupling $\beta_c$ by performing a linear 
two-parameter fit, equ.(\ref{eq:bc}), using 
the maxima of $C$ and $\chi$ along with derivatives
of the three variants of $U_2$ and $U_4$ and derivatives
and logarithmic derivatives of the magnetization as estimators of
the pseudocritical
points. This gives the eleven estimators which are shown in Table~2.
A  global estimate is
again extracted by performing both a straightforward
average and an error weighted average.
It is noteworthy that the estimated critical couplings are compatible with
those found in simulations of the models 
coupled to 2D quantum gravity \footnote{On corresponding
ensembles of graphs without tadpoles and self-energy bubbles.},
i.e., with {\it annealed} rather than quenched connectivity disorder. 
This was already remarked on in \cite{9,10} where non-linear
three parameter fits were employed.
                                                                     
 \begin{table}[htbp] 
\centering
\caption[{\em Nothing.}]
 {{\em Fit results for the pseudocritical couplings $\beta_c$.}}  
\vspace{3ex}
   \begin{tabular}{|r|l|l|l|}
 \hline\hline
 \multicolumn{1}{|c|}{at $ \beta_{\rm max}$ of} &
 \multicolumn{1}{c|}{$q=2$}   &
 \multicolumn{1}{c|}{$q=4$}   &
 \multicolumn{1}{c|}{$q=10$}   \\
 \hline
 $C$ &               1.539(5)  &  1.836(1)  &  2.244(1)   \\
 $\chi$ &            1.562(6)  &  1.834(2)  &  2.246(1)   \\
 $dU_2^{(1)}/d \beta$ &   1.551(5)  &  1.831(1)  &  2.244(1)   \\
 $dU_2^{(2)}/d \beta$ &   1.550(5)  &  1.834(1)  &  2.242(1)   \\
 $dU_2^{(3)}/d \beta$ &   1.550(5)  &  1.831(2)  &  2.243(1)   \\
 $dU_4^{(1)}/d \beta$ &   1.558(7)  &  1.832(2)  &  2.244(1)   \\
 $dU_4^{(2)}/d \beta$ &   1.569(4)  &  1.841(2)  &  2.240(2)   \\
 $dU_4^{(3)}/d \beta$ &   1.562(5)  &  1.834(2)  &  2.242(1)   \\
 $d \langle m \rangle/d \beta$ &    1.561(3)&  1.834(1)  &  2.245(1)  \\
 $d\log\langle m \rangle/d \beta$   & 1.562(5)  & 1.837(2) &  2.243(1) \\
 $d\log\langle m^2 \rangle/d \beta$ &  1.570(5) & 1.838(1) &  2.243(1) \\
 \hline
 {\bf average} &           {\bf 1.558(3)}  &  {\bf 1.835(1)}  &  {\bf 2.244(1)}   \\
 {\bf weighted av.} &      {\bf 1.558(3)}  &  {\bf 1.835(1)}  &  {\bf 2.244(1)}   \\
 \hline\hline
 \end{tabular}
 \end{table}

\begin{table}[htbp]
\centering
\caption[{\em Nothing.}]
 {{\em Fit results for the critical exponent $\alpha /\nu D$ for $q=10$.}} 
\vspace{3ex}
   \begin{tabular}{|l|l|}
 \hline\hline
 \multicolumn{1}{|c|}{at $\beta_{\rm max}$ of} &
 \multicolumn{1}{c|}{$q=10$}   \\
 \hline
 $C$ &                                   0.21(1)   \\
 $\chi$ &                                0.21(2)   \\
 $dU_2^{(1)}/d \beta$ &                       0.22(1)   \\
 $dU_2^{(2)}/d \beta$ &                       0.21(1)   \\
 $dU_2^{(3)}/d \beta$ &                       0.22(1) \\
 $dU_4^{(1)}/d \beta$ &                       0.21(1)  \\
 $dU_4^{(2)}/d \beta$ &                       0.18(2)   \\
 $dU_4^{(3)}/d \beta$ &                       0.21(1)   \\
 $d \langle m \rangle/d \beta$ &         0.21(1)   \\
 $d\log\langle m \rangle/d \beta$ &      0.21(1)   \\
 $d\log\langle m^2 \rangle/d \beta$ &    0.22(1)    \\
 \hline
{\bf  average} &                             {\bf   0.21(1)}   \\
{\bf  weighted av.} &                        {\bf   0.21(1)}   \\
 \hline\hline
$q= 4 \; Quenched$ &   0.177\dots\\
\hline
$q= 4 \; KPZ$      &   0 \\
\hline
$q= 4 \; Regular$  &   0.5 \\ 
 \hline\hline
 \end{tabular}
 \end{table}

The crossing points of the various definitions of
the fourth-order cumulant in eqs.~(\ref{eq:U4_1}) and (\ref{eq:U4_23})
provide further estimates of $\beta_c$. The error bars
from these measurements are much larger than those
in Table~2 due to the spread of crossings for different system
sizes, but the estimates from $U_4^{(1)}$, $U_4^{(2)}$, and $U_4^{(3)}$
are all consistent, giving 
$\beta_c=1.58(2)$ for $q=2$,
$\beta_c=1.85(2)$ for $q=4$, and
$\beta_c =2.244(4)$ for $q=10$.
The rather wide spread in the estimated $\beta_c$ is reflected
by a similar spread in the values of $U_4$ at $\beta_c$, usually denoted
by $U^*$. The variant $U_4^{(2)}$ appears to be the best behaved for all 
$q$ giving $U^* \approx 0.55(3)$.

Moving on to $\alpha / \nu D$, the Fisher hyperscaling relation
in the form
\begin{equation}
{\alpha \over \nu D} = {2 \over \nu D} - 1
\end{equation}
gives $\alpha / \nu D = -0.32(1),\;  -0.16(1)$, and  $0.16(1)$ for
$q=2$, 4, and 10, respectively, from the $1 / \nu D$ values in
Table~1, whereas the quenched predictions are $\alpha / \nu D = -0.303\dots$
and $0.177\dots$
for $q=2$ and $4$. Direct non-linear fits to $\alpha / \nu D$ using the scaling 
form $C = C_{reg} + C_1 N^{\alpha / \nu D}$ for
the maxima of the specific heat are 
unstable for $q=2,4$ but give  $\alpha / \nu D  = 0.22(7)$ for $q=10$.
The constant in this fit is consistent with zero, so we carried out a
log-log fit at all of the pseudo-critical points as shown in Table~3,
giving a final average of  $\alpha / \nu D  = 0.21(1)$ for $q=10$.
Two representative series of points for the 
specific heat evaluated at its own maximum and the maximum
of the susceptibility on the different graph sizes
are shown in Fig.~3.
This independent measurement of $\alpha / \nu D$ when $q=10$ is thus
still in reasonable agreement with the theoretical quenched 
exponent values for $q=4$.

In Tables~4,~5, and 6 we list the measured values of the magnetic
exponents $\gamma / \nu D$, $\beta / \nu D$ and $(1 - \beta ) / \nu D$
for $q=2,4$, and $10$. We have adopted a similar approach to that used 
in the estimation of $1 / \nu D$. 
Data from all but the smallest graph size was included in the fits,
and the stability of the fits checked against the deletion of 
the next smallest sizes, proving in all the listed cases to be reasonable.
The appropriate 
scaling relations in equ.~(\ref{eq:fss_m})
are evaluated at all the pseudo-critical points 
reported in the tables and a final averaged
value and error weighted average calculated. 
For the blank entries in the tables no stable fit proved possible.
In the tables the number of terms
used in the averages for each exponent are listed in the average row,
and various theoretical predictions for the exponents are listed
for comparison. Since it is already clear that the 
$q=10$ measurements bear a considerable similarity to the quenched
$q=4$ predictions we have repeated the $q=4$ theoretical values
in the $q=10$ table. The exponent ratio $\gamma / \nu D$ is evaluated from 
the scaling
of the susceptibility $\chi$, $\beta / \nu D$ from the scaling of 
the derivative of the magnetisation and $(1 - \beta ) / \nu D$
from the scaling of the logarithmic derivative.

The scaling relation
\begin{equation}
{\gamma \over \nu D} = 1 - 2 { \beta \over \nu D}
\label{eq:scal2}
\end{equation}
relates the two exponents in these tables. The directly measured values
are all in reasonable agreement with those derived by using either
exponent as input in this scaling relation. There is an
accidental equality between the theoretical quenched values of
${\gamma / \nu D}$ (and hence via equ.(\ref{eq:scal2}) of $ \beta / \nu D$)
for the Ising and 4-state Potts models which  is not, however, reflected in the estimates.
There is a steady decrease in ${\gamma / \nu D}$
and corresponding increase in $\beta / \nu D$ as $q$ is increased,
so the $q=4$ and $q=10$ measurements are clearly different from those for $q=2$.
There is still a striking agreement between the quenched $q=4$
predictions and the measurements at $q=10$. It is also clear
that the
$q=4$ measurements are in definite disagreement with the quenched $q=4$ predictions.

In Fig.~4 we plot the data points and fits for the susceptibility
in the Ising model. The individual points for $\chi$ evaluated
at its own maximum, the maximum of the specific heat, the maxima
of the derivatives of the second and fourth order cumulants 
and the maximum of the derivative of the magnetisation are
shown explicitly \footnote{We have shown only the $U_{2,4}^{(1)}$
cumulant results to avoid cluttering the graph, the alternative
definitions give effectively identical results. $\chi$ evaluated at the maxima
of the derivatives
of the log of the magnetisation and the modulus squared of the magnetisation
has also been dropped for clarity. The graphs for $q=4$ and $q=10$ show similar
features and are not reproduced here.}. The fits from
the different series are all in good agreement, as indicated by the tables. 
We also plot $\chi$ evaluated
at its own maximum for $q=2$, 4, and 10 in Fig.~5 in order to show the general
trend in the exponent.
%
                     \section{Conclusions}
%
The qualitative conclusions of our extensive analyses are quite clear:
the quenched bond disorder of the $\Phi^3$ graphs {\it does}
alter the exponents of models which already possess a 
continuous transition on a regular lattice, as well as softening
the first-order transition of the $q=10$ model to a continuous transition.
A quenched ensemble of $\Phi^3$ graphs 
with connectivity disorder thus shares many of the features
of a system with quenched bond disorder. As noted in the introduction,
other ensembles with quenched connectivity disorder
such as Poisonnian random lattices do not -- being more
akin to regular lattices.
 
At a quantitative level, however, the current batch of simulations pose 
rather more questions than they answer:
the working hypothesis of the veracity of
the quenched exponents is 
at best only weakly supported by the results. 
For the Ising ($q=2$) model, the estimated value of $1/ \nu D$, $0.34(1)$, is 
consistent with both the quenched and KPZ predictions. 
Although the estimates for the magnetic exponents 
$[\gamma / \nu D, ~\beta / \nu D, ~( 1 - \beta ) / \nu D] \sim [0.79(1), ~0.11(1), ~0.26(1) ]$ 
are closer to the quenched $[0.7094, ~0.1452, ~0.2033]$ than the Onsager $[0.875, ~0.0625, ~0.4375]$
or KPZ $[0.666.., ~0.166..., ~0.166...]$
values, any agreement is less than convincing. In mitigation, it is fair to point out that we have struggled in the past
to obtain good estimates of the magnetic KPZ exponents on dynamical $\Phi^3$ graphs
without tadpoles and self-energy bubbles \cite{bj}, 
which is the class of graph we have used in the simulations here. This is
probably due to large corrections to scaling, since including degenerate
graphs (self-energy bubbles and tadpoles) appears to give a faster approach to the continuum 
limit \cite{amb_thor}.

Taken {\it en masse} the estimated exponents of the $q=4$ model,
$[1/ \nu D, ~\gamma / \nu D, ~\beta / \nu D, ~( 1 - \beta ) / \nu D ] 
\sim [0.42(1), ~0.75(1), ~0.12(1), ~0.34(1) ]$
fit neither the quenched [0.589\dots, ~0.7094\dots, ~0.1452\dots, ~0.4433\dots], KPZ 
[0.5, ~0.5, ~0.25, ~0.25] nor regular lattice [0.75, ~0.875, ~0.0625, ~0.6875]
predictions, although one could argue that $\gamma / \nu D \sim 0.75(1)$  on its own is
actually closer to the quenched prediction than the $q=2$ model. 
It is possible that $q=4$, which is subject to logarithmic corrections
in both the regular lattice and KPZ cases, may require similar treatment in the quenched case
but without more input on the form of these corrections fitting would be a futile exercise.

The $q=10$ model, on the other hand,
provides us with a set of estimated exponents $[1/ \nu D, ~\gamma / \nu D, ~\beta / \nu D, ~( 1 - \beta ) / \nu D] 
\sim [0.58(1), ~0.71(1), ~0.12(1), ~0.43(2) ]$
which match the predicted
quenched $q=4$ exponents $[0.589.., ~0.7094.., ~0.1452.., ~0.4433.. ]$ extremely well. 
These estimated $q=10$ exponents
(and theoretical $q=4$ values)
hence violate the bound $\nu D > 2$
of \cite{18a}, which merits an explanation in its own right.
 
It is difficult to quibble with the thermal statistics 
from the very long time series in the current batch of simulations,
but one might find fault with the relatively modest number
of replicas used in the disorder averaging. The replica to replica
variation of the measured quantities does not, however, appear
to have significantly skewed the measured values and error estimates.
Looked at without theoretical prejudice the measured exponents
suggest a slow variation with $q$ that is akin to that observed
in the quenched bond disorder simulations of \cite{4},
rather than values which stick at $q=4$ and change no further with increasing $q$.
It is thus possible that the agreement of the measured $q=10$ exponents with
the predicted $q=4$ values is accidental. 
A simple way to settle this issue would
be to simulate other values of $q$ to investigate the variation, if any,
of the exponents, particularly for $q>4$.

There is a second parameter that one can vary in such simulations,
namely the central charge used in generating the graphs of 
the quenched ensemble.
Since the 
partition function $Z_N$ obtained on integrating out $d$
scalar fields on $\Phi^3$ graphs with $N$ vertices
is 
\begin{equation}
 Z_N \;=\; \sum_{G\in{\cal G}(A)} ({\rm det} \; C_G)^{-d/2} \;,
\label{eq:gud}
\end{equation}
where $\cal G (A)$ is the
class of graph being summed over
and $C_G$ is the adjacency matrix of the graph $G$,
\begin{equation}
 C_G \;=\; \left \{
 \begin{array}{ll}
  q_i  & \qquad \text{if $i=j$,} \\
  -n_{ij}  & \qquad \text{if $i$ and $j$ are adjacent,} \\
  0     & \qquad \text{otherwise,}
 \end{array}
 \right .
\end{equation}
one can use equ.(\ref{eq:gud}) to generate an ensemble of graphs
to which one can associate a central charge $d$. 

Quenched
simulations may then be carried out on this ensemble
rather than the $d=c=0$ graphs used here. 
The appropriate $d$ can be substituted
into the $KPZ$ formula to obtain predictions for the 
exponents in this case. This brings one to another
puzzle: in \cite{19} very good agreement was found between the 
predicted exponents for the Ising model on a 
quenched ensemble of graphs with $d = -5$ and measurements.
In this light, it is surprising that the agreement 
here for the Ising model on much larger $d=0$ graphs with better
statistics is poorer. One could speculate that the scaling
behaviour of the models improved as they became more ``classical'',
i.e. as the effects of gravity were switched off ($d \rightarrow - \infty$).

Various other aspects of the behaviour of spin models on quenched random 
gravity graphs 
that have only been touched on here  merit further investigation.
The clear evidence of non-self-averaging behaviour for all $q$ and
the autocorrelation scaling techniques used to quantify it are described in more detail in a 
companion paper \cite{jj}. Similarly, the multifractal scaling of spin correlation functions
can also be investigated. Finally, as we have already noted, further simulations for other $q$
(and also $d$) values would help to determine whether the quenched exponents
were correctly describing the behaviour of the models and cast further light on the
remarkable (accidental?) agreement between $q=10$ measurements and the predicted $q=4$ 
exponents.

In summary, spin models
on $\Phi^3$ random graphs offer a useful
framework for the exploration of quenched disorder
and may even offer some advantages over bond disordered models
given the availability of various exact, rather than perturbative,
predictions for exponents. The results described here suggest numerous 
avenues for future work.

%
%
\section*{Acknowledgements}
%
%
DJ was partially supported by a Royal Society of
Edinburgh/SOEID Support Research Fellowship.
WJ acknowledges partial support by the German-Israel-Foundation (GIF) under
contract No.\ I-0438-145.07/95.
The collaborative work of DJ and WJ was funded by ARC grant
313-ARC-XII-98/41. The numerical simulations were performed on a T3D parallel
computer of Zuse-Zentrum f\"ur Informationswissenschaften Berlin (ZIB)
under grant No.~bvpf01.
\bigskip
%

%
%
\clearpage \newpage
%
\begin{table}[t]
\centering
\caption[{\em Nothing.}]
 {{\em q=2 Potts (Ising) fit results for the critical exponents $\gamma/\nu D$,
 $\beta/\nu D$, and $(1-\beta)/\nu D$. 
 }}
\vspace{3ex}
\begin{tabular}{|l|ll|ll|ll|}
\hline\hline
\multicolumn{1}{|c|}{at $\beta_{\rm max}$ of}  &
\multicolumn{1}{c}{$\gamma/\nu D$}           &
\multicolumn{1}{c|}{$Q$}                   &
\multicolumn{1}{c}{$\beta/\nu D$}            &
\multicolumn{1}{c|}{$Q$}                   &
\multicolumn{1}{c}{$(1-\beta)/\nu D$}        &
\multicolumn{1}{c|}{$Q$}                   \\
\hline
 $C$                                              & 0.75(1)  & 0.14  & 0.14(2)  & 0.24 & 0.23(2) &0.40  \\
 $\chi$                                           & 0.79(1)  & 0.13  & 0.08(1)  & 0.21 & 0.26(1) &0.78  \\
 $dU_4^{(1)}/d \beta$                             & 0.79(2)  & 0.66  & 0.12(1)  & 0.42 & 0.26(2) &0.33  \\
 $dU_4^{(2)}/d \beta$                             & --       & --    & --       & --   & --      & --   \\
 $dU_4^{(3)}/d \beta$                             & 0.80(2)  & 0.46  & 0.10(1)  & 0.14 & 0.28(1) &0.13  \\
 $dU_2^{(1)}/d \beta$                             & 0.77(1)  & 0.75  & 0.12(1)  & 0.82 & 0.25(1) &0.76   \\
 $dU_2^{(2)}/d \beta$                             & 0.77(1)  & 0.73  & 0.12(1)  & 0.64 & 0.25(1) &0.63   \\
 $dU_2^{(3)}/d \beta$                             & 0.77(1)  & 0.78  & 0.12(1)  & 0.84 & 0.25(1) &0.62   \\
 $d [\langle |m| \rangle]_{\rm av}/d \beta$       & --       & --    & 0.09(1)  & 0.38 & 0.26(1) &0.80  \\
 $d\ln[\langle |m| \rangle]_{\rm av}/d \beta$     & 0.80(1)  & 0.55  & 0.10(1)  & 0.36 & 0.28(1) &0.37  \\
 $d\ln[\langle m^2 \rangle]_{\rm av}/d \beta$     & 0.83(2)  & 0.15  & --       & --   & 0.30(1) &0.32  \\
\hline\hline
 {\bf average (9,9,10)}  &   {\bf 0.79(1)}     & {}  & {\bf 0.11(1)}  & {} & {\bf 0.26(1)}  & {} \\
 \hline
 {\bf weighted av.} &   {\bf 0.78(1)}      & {}  & {\bf 0.10(1)}  & {} & {\bf 0.26(1)}      & {}\\
 \hline\hline                                       
$Quenched$ &   0.7094\dots & {}  & 0.1452\dots & {} & 0.2033\dots & {} \\
\hline
$KPZ$      &   0.6666\dots & {}  & 0.1666\dots & {} & 0.1666\dots & {} \\
\hline
$Regular$  &   0.875     & {}  & 0.0625    & {} & 0.4375    & {} \\
\hline\hline                                            

\end{tabular}
\label{tab:q2}
\end{table}

%
\begin{table}[htbp]
\centering
\caption[{\em Nothing.}]
 {{\em q=4 Potts fit results for the critical exponents $\gamma/\nu D$,
 $\beta/\nu D$, and $(1-\beta)/\nu D$.
 }}
\vspace{3ex}
\begin{tabular}{|l|ll|ll|ll|}
\hline\hline
\multicolumn{1}{|c|}{at $\beta_{\rm max}$ of}  &
\multicolumn{1}{c}{$\gamma/\nu D$}           &   
\multicolumn{1}{c|}{$Q$}                   &
\multicolumn{1}{c}{$\beta/\nu D$}            &   
\multicolumn{1}{c|}{$Q$}                   &
\multicolumn{1}{c}{$(1-\beta)/\nu D$}        &   
\multicolumn{1}{c|}{$Q$}                   \\
\hline
 $C$                                              & --       & --    &   --     & --   & --      & --    \\
 $\chi$                                           & 0.75(1)  & 0.13  &  --      & --   & 0.34(2) & 0.61 \\
 $dU_4^{(1)}/d \beta$                             & --       & --    &  --      & --   & 0.33(1) & 0.37 \\
 $dU_4^{(2)}/d \beta$                             & --       & --    &   --     &  -- & --      & --    \\
 $dU_4^{(3)}/d \beta$                             & --       & --    &   --     &  -- & --      & --    \\
 $dU_2^{(1)}/d \beta$                             & 0.75(1)  & 0.82  & 0.12(1)  & 0.50 & 0.33(1) & 0.70   \\
 $dU_2^{(2)}/d \beta$                             & 0.77(1)  & 0.42  & --       & --   & 0.34(1) & 0.41   \\
 $dU_2^{(3)}/d \beta$                             & 0.75(1)  & 0.80  & 0.13(1)  & 0.26 & 0.33(1) & 0.66   \\
 $d [\langle |m| \rangle]_{\rm av}/d \beta$       & 0.75(1)  & 0.44  & 0.10(1)  & 0.99 & 0.34(1) & 0.73  \\
 $d\ln[\langle |m| \rangle]_{\rm av}/d \beta$     & --       & --    & --       & --   & --      & --    \\
 $d\ln[\langle m^2 \rangle]_{\rm av}/d \beta$     & --       & --    & --       & --   & --      & --    \\
 \hline\hline
{\bf  average (5,3,6)}  &   {\bf 0.75(1)}     & {}  & {\bf 0.12(1)} & {} & {\bf 0.34(1)}  & {} \\
 \hline
{\bf  weighted av.} &   {\bf 0.75(1)}      & {}  & {\bf 0.11(1)}  & {} & {\bf 0.34(1)}   & {}\\  
 \hline\hline
$Quenched$ &   0.7094\dots     & {}  & 0.1452\dots & {} & 0.4433\dots  & {} \\
\hline
$KPZ$      &   0.5          & {}  & 0.25     & {} & 0.25      & {}\\
\hline
$Regular$  &   0.875        & {}  & 0.0625   & {} & 0.6875    & {}\\
\hline\hline                                       
\end{tabular}
\label{tab:q4}
\end{table}

%
\begin{table}[t]
\centering
\caption[{\em Nothing.}]
 {{\em q=10 Potts fit results for the critical exponents $\gamma/\nu D$,
 $\beta/\nu D$, and $(1-\beta)/\nu D$.
 }}
\vspace{3ex}
\begin{tabular}{|l|ll|ll|ll|}
\hline\hline
\multicolumn{1}{|c|}{at $\beta_{\rm max}$ of}  &
\multicolumn{1}{c}{$\gamma/\nu D$}           &   
\multicolumn{1}{c|}{$Q$}                   &
\multicolumn{1}{c}{$\beta/\nu D$}            &   
\multicolumn{1}{c|}{$Q$}                   &
\multicolumn{1}{c}{$(1-\beta)/\nu D$}        &   
\multicolumn{1}{c|}{$Q$}                   \\
\hline
 $C$                                              & 0.71(2)  & 0.13  & --      & --   & --       & --   \\
 $\chi$                                           & 0.72(1)  & 0.31  & 0.10(1) & 0.39 & 0.43(2)  & 0.10 \\
 $dU_4^{(1)}/d \beta$                             & 0.73(2)  & 0.72  & 0.11(1) & 0.23 & 0.43(2)  & 0.53 \\
 $dU_4^{(2)}/d \beta$                             & --       & --    & --      & --   & --       & --   \\
 $dU_4^{(3)}/d \beta$                             & 0.69(2)  & 0.30  & --      & --   & 0.41(2)  & 0.22  \\
 $dU_2^{(1)}/d \beta$                             & 0.72(2)  & 0.61  & 0.12(1) & 0.32 & 0.43(2)  & 0.19  \\
 $dU_2^{(2)}/d \beta$                             & --       & --    & --      & --   & 0.42(2)  & 0.16   \\
 $dU_2^{(3)}/d \beta$                             & 0.71(2)  & 0.37  & 0.14(1) & 0.29 & 0.43(2)  & 0.12  \\
 $d [\langle |m| \rangle]_{\rm av}/d \beta$       & 0.72(2)  & 0.25  & 0.11(1) & 0.10 & 0.43(2)  & 0.11 \\
 $d\ln[\langle |m| \rangle]_{\rm av}/d \beta$     & 0.71(2)  & 0.10  & --      & --   & 0.43(2)  & 0.10 \\
 $d\ln[\langle m^2 \rangle]_{\rm av}/d \beta$     & --       & --    & --      & --   & 0.43(2)  & 0.11  \\
 \hline\hline
{\bf  average (8,5,9)}  &   {\bf 0.71(1) }    & {}  & {\bf 0.12(1)} & {} & {\bf 0.43(2) } & {} \\
 \hline
{\bf  weighted av.} &   {\bf 0.71(1)}      & {}  & {\bf 0.12(1)}  & {} & {\bf 0.43(2) }  & {}\\
 \hline\hline                                                            
$q=4 \; \; Quenched$ &   0.7094\dots     & {}  & 0.1452\dots & {} & 0.4433\dots  & {} \\
\hline
$q=4 \; \; KPZ$      &   0.5          & {}  & 0.25     & {} & 0.25      & {}\\
\hline
$q=4 \; \; Regular$  &   0.875        & {}  & 0.0625   & {} & 0.6875    & {}\\
\hline\hline                                       
\end{tabular}
\label{tab:q10}
\end{table}

\vfill
%
\clearpage \newpage
\begin{figure}[htb]
\vskip 20.0truecm
\includegraphics{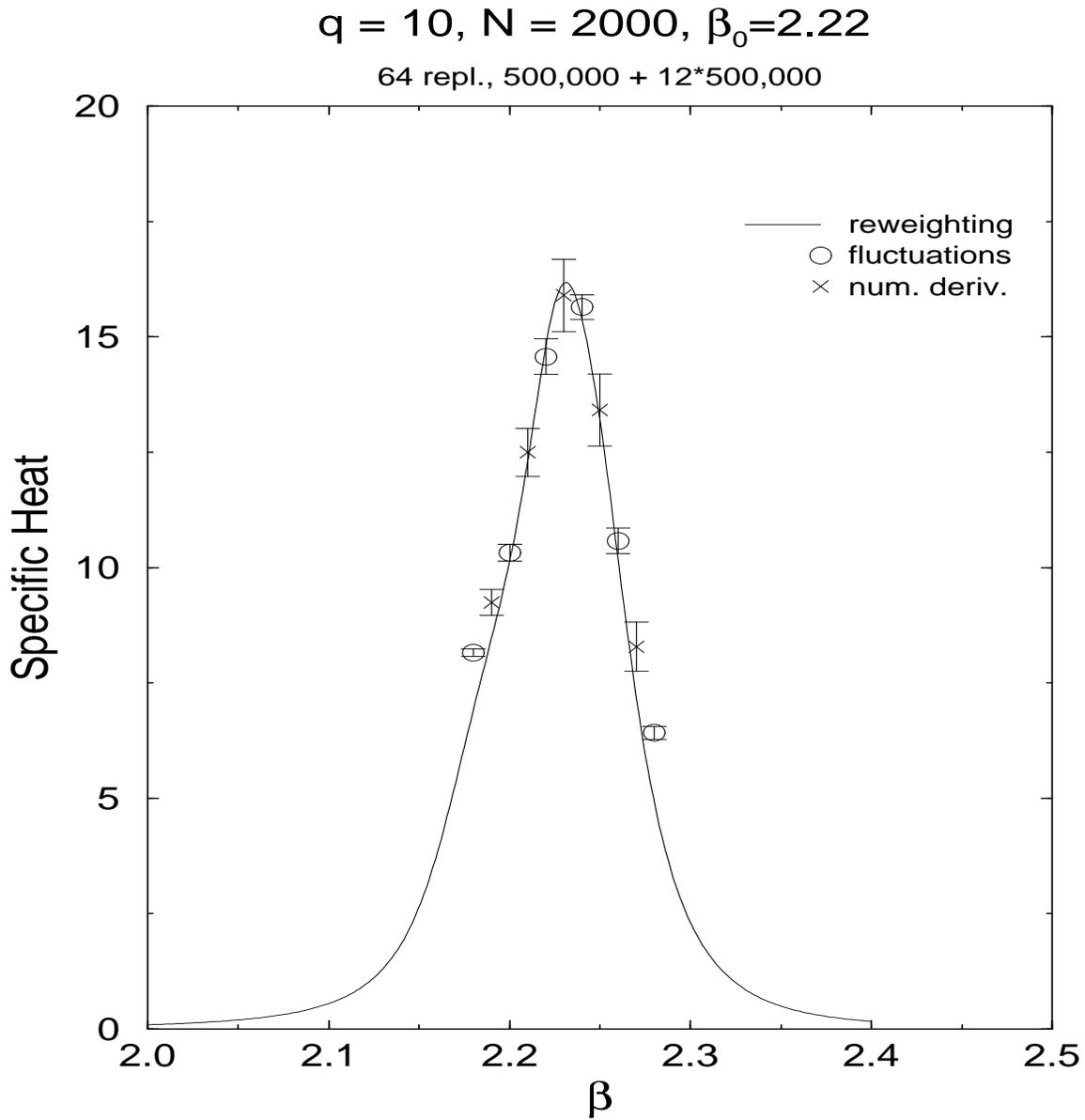}
\caption[]{\label{fig1} The specific heat calculated from reweighting
about $\beta_0=2.22$,
fluctuations and numerical differentiation of the energy for
$q=10$ and $N=2000$ (which is typical).
}
\end{figure}        
\clearpage \newpage

\begin{figure}[htb]
\vskip 20.0truecm
\includegraphics{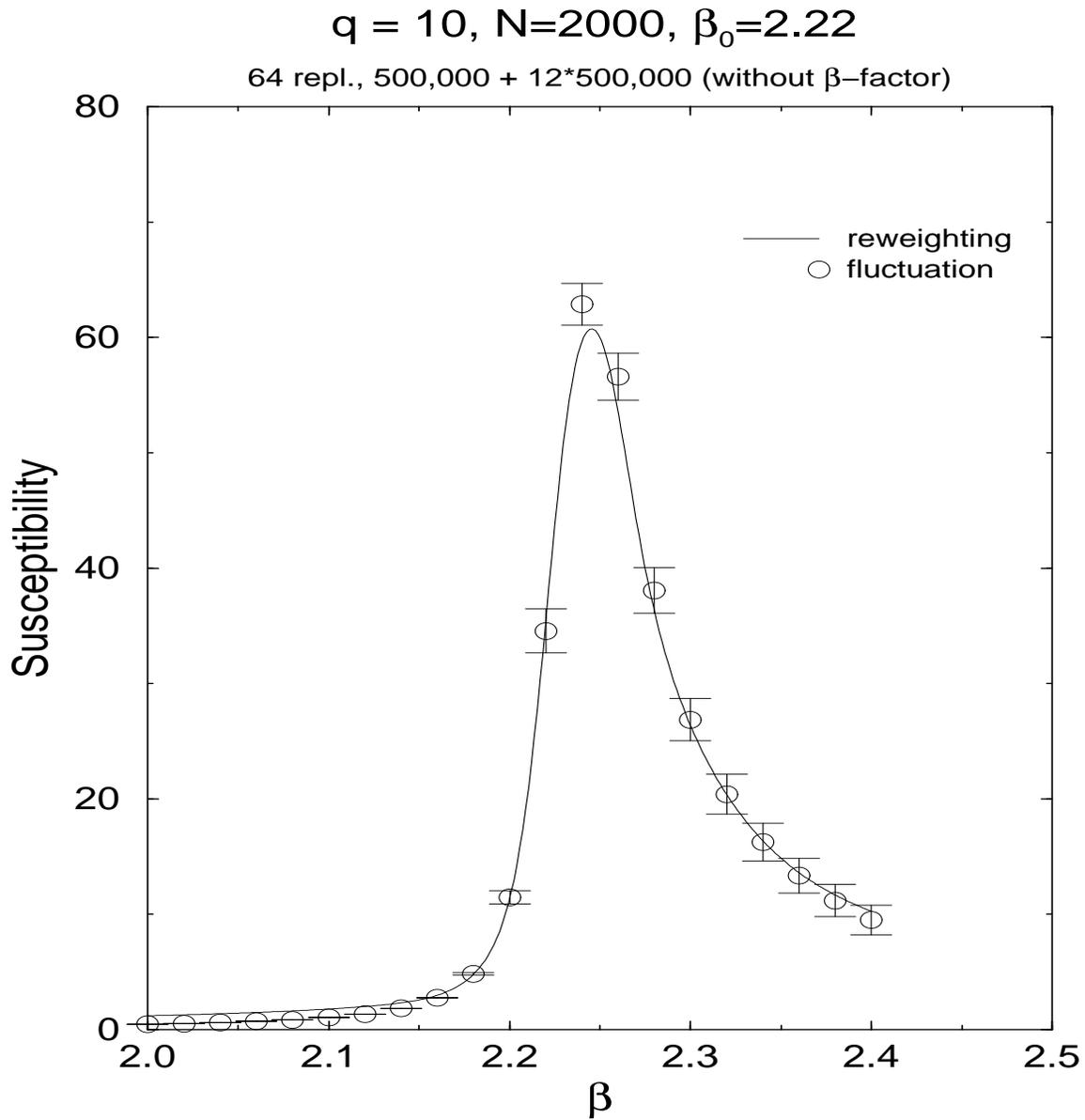}
\caption[]{\label{fig2} The susceptibility calculated from reweighting
about $\beta_0=2.22$ and from
fluctuations for $q=10$ and $N=2000$.
}
\end{figure}                                     

\clearpage \newpage

\begin{figure}[htb]
\vskip 20.0truecm
\includegraphics{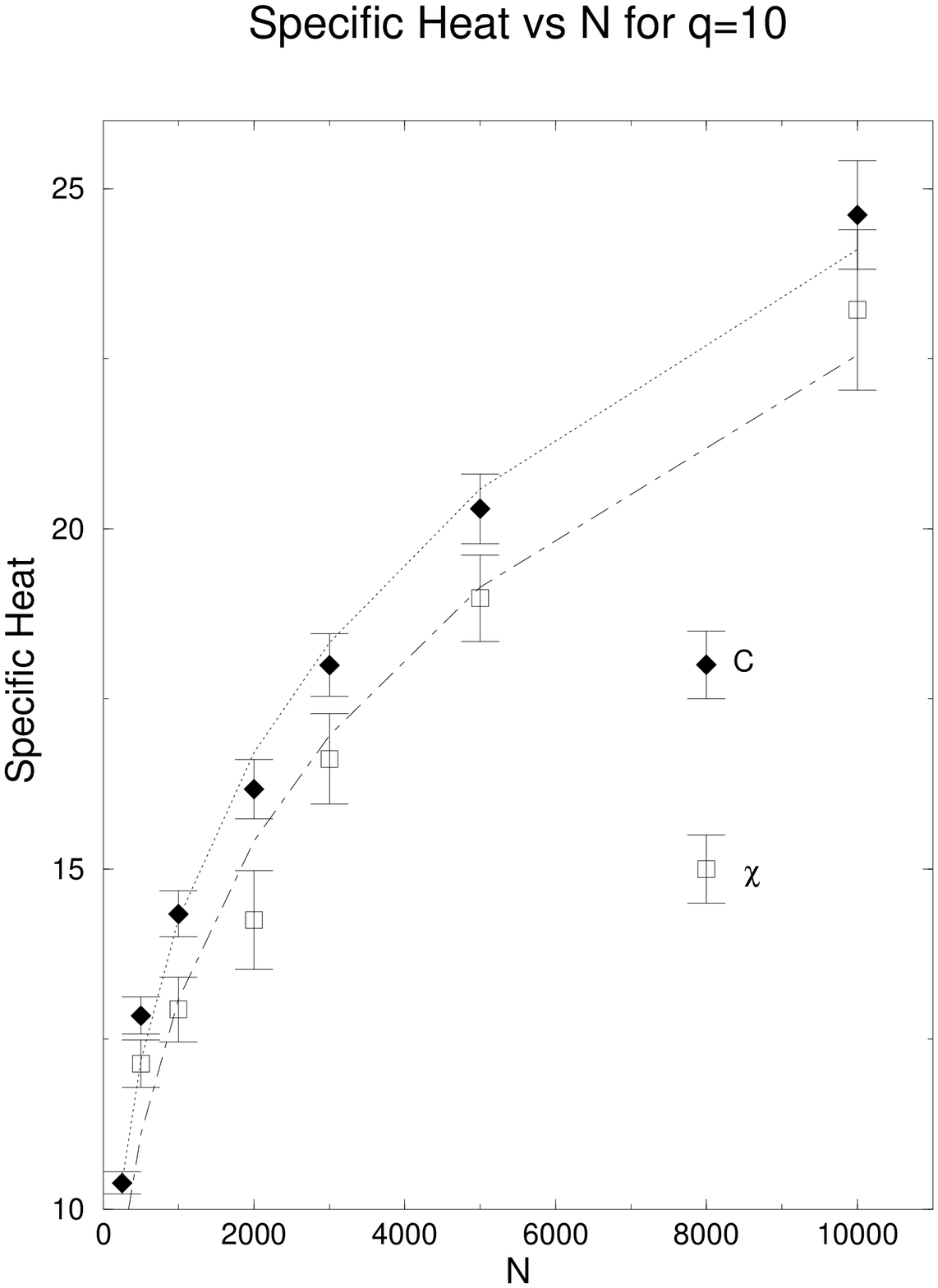}
\caption[]{\label{fig5} Two representative fits 
(from the eleven used) for the scaling
of the specific heat evaluated at its own maximum and at the maximum of
the susceptibility $\chi$.
}

\end{figure}           
\clearpage \newpage %
 
\begin{figure}[htb] 
\vskip 20.0truecm 
\includegraphics{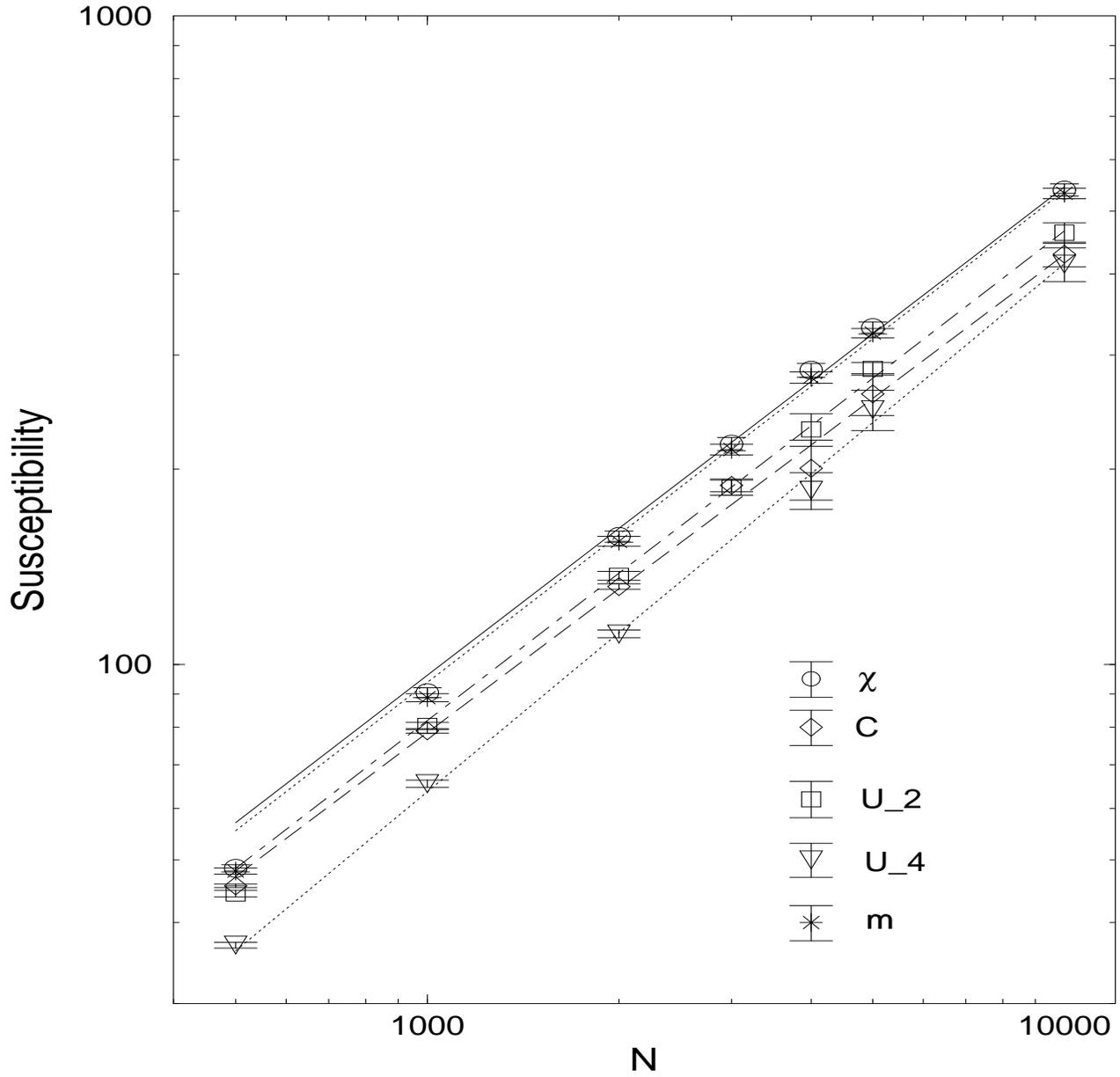} 
\caption[]{\label{fig6} Fits to $\chi$  when $q=2$ to obtain $\gamma / \nu D$ at
the maxima of $\chi$, $C$, $d U_2^{(1)} / d \beta$, $d U_4^{(1)} / d \beta$
and $d m / d \beta$ are shown on a log-log scale.}
 
\end{figure}

\clearpage \newpage %
 
\begin{figure}[htb]
\vskip 20.0truecm
\includegraphics{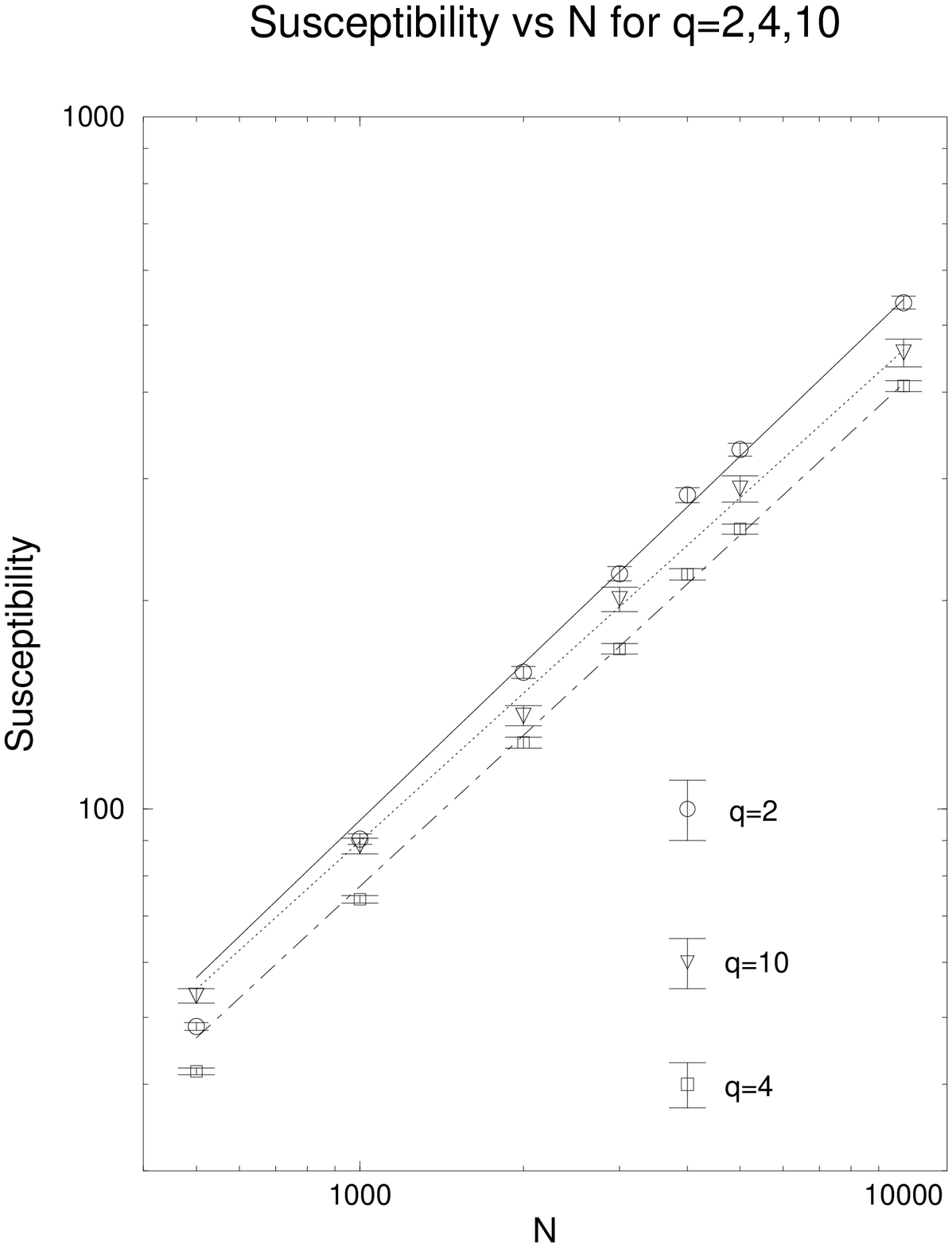}
\caption[]{\label{fig7} Fits to $\chi$  at its own maximum 
for $q=2,4$, and $10$.}
 
\end{figure}                                         
\end{document}